\documentclass[aps,prl,reprint,groupedaddress]{revtex4-1}

\usepackage[utf8x]{inputenc}
\usepackage{amsfonts,amssymb}
\usepackage{graphicx}
\usepackage{hyperref}
\usepackage{caption}
\usepackage{subcaption}
\usepackage{amsmath}
\usepackage{amssymb}
\usepackage{amsthm}
\usepackage{color}
\usepackage{dsfont}

\newcommand{\R}{\mathbb{R}}

\newcommand{\pP}{\mathbb{P}}

\begin{document}


\title{Cycle flow based module detection in directed recurrence networks}


\author{Ralf Banisch}
\email[]{ralf.banisch@fu-berlin.de}
\affiliation{Institut f\"ur Mathematik und Informatik, Freie Universität Berlin}


\author{Nata\v{s}a Djurdjevac Conrad}
\email[]{djurdjev@fu-berlin.de}
\affiliation{Institut f\"ur Mathematik und Informatik, Freie Universität Berlin}
\altaffiliation{Zuse Institute Berlin}

\date{9 January 2015}

\begin{abstract}
We present a new cycle flow based method for finding fuzzy partitions of weighted directed networks coming from time series data. We show that this method overcomes essential problems of most existing clustering approaches, which tend to ignore important directional information by considering only one-step, one-directional node connections. Our method introduces a novel measure of communication between nodes using multi-step, bidirectional transitions encoded by a cycle decomposition of the probability flow. Symmetric properties of this measure enable us to construct an undirected graph that captures information flow of the original graph seen by the data and apply clustering methods designed for undirected graphs. Finally, we demonstrate our algorithm by analyzing earthquake time series data, which naturally induce (time-)directed networks.\\
This article has been published originally in EPL, DOI: \href{http://iopscience.iop.org/article/10.1209/0295-5075/108/68008?fromSearchPage=true}{10.1209/0295-5075/108/68008}. This version differs from the published version by minor formatting details.
\end{abstract}



\maketitle


\section{Introduction}

Real-world data is often analyzed by constructing appropriate networks from it. Inferring properties of such data is thus closely related to studying the associated networks. For example, the presence of communities or {\bf modules} in the network often indicates relevant structure in the data. Module identification is a very well studied area of research in network theory, and different approaches have been proposed \cite{Newman04,NewmanGirvan1,SVDongen00,DdWb05}, see \cite{Newman2003b,Fortunato2010} for an exhaustive review. The goal is to find a clustering of the node set $V$ which is provided by $m$ affiliation functions $q_i: V \rightarrow [0,1]$. In this article, we are interested in clustering networks constructed from data with the following properties: (i) the data is a time-ordered series $\mathbf{x}_{[1,T]} = \{x_1,\ldots, x_T\}$ in an observational space $\Omega\subset \R^d$. (ii) The data is not perfectly structured. As a result, the network won't be either. A perfectly structured network gives rise to crisp affiliation functions, i.e. every node belongs to exactly one module. Imperfectly structured networks give rise to fuzzy affiliation functions indicating uncertainty in the clustering. These are typical properties of data coming from real-world systems, for example metastable time series in molecular dynamics or meteorological data, consisting of metastable sets and states in a transition region.\\
Network based time series analysis has gained a lot of attention in the last years, which resulted in different methods for constructing networks from time series data, see \cite{Donner2010} for a review. Here we will adopt an approach based on the well established framework of symbolic dynamics\cite{Robinson1999}: partition $\Omega$ into $N$ disjoint sets $\{S_1,\ldots,S_N\}$, identify the graph nodes with the sets $S_i$ and characterize the edges by the transition probabilities
\begin{equation}
 P_{xy} = \pP(x_{t+1}\in S_y | x_t \in S_x).\label{eq:Tmatrix}
\end{equation}
This leads to a so-called recurrence network, i.e. a weighted, directed network in which directions reflect the time-ordering, and we account for (ii) by searching for a fuzzy clustering of the network. Note that we do not restrict our approach only to time series data, but we can consider networks directly. In this case, a random walk process is defined on the network and its realization is used as an input for the algorithm.\\
Most methods for community detection are designed for undirected networks and purely density-based, i.e. they seek to maximize the density of links within modules and minimize the density of links between modules. The most prominent example is modularity optimization \cite{Newman04,NewmanGirvan1} and its generalizations \cite{Clauset2004,Arenas2007,Kim2010}. However, modularity based methods consider only one-directional, one-step transitions and as such they are blind to the directional structures in the network, see \cite{Fortunato2009}. Dynamics-based multistep community detection algorithms, like Markov stability \cite{Lambiotte09} and Infomap \cite{Delvenne2010}, can deal with directed networks in a natural way. But these algorithms only produce hard partitionings, i.e. every node is assigned to exactly one module.\\
In this paper we will use the Markov State Model (MSM) clustering method \cite{Djurdjevac2011,SarichNet11}, a dynamics-based fuzzy clustering algorithm for finding multiscale clusters. Since MSM relies crucially on the time-reversibility of the transition matrix (\ref{eq:Tmatrix}), which restricts the algorithm to undirected networks, as a main result of this paper we will present an algorithm to construct a reversible transition matrix (\ref{eq:CMSM}) from the data $\mathbf{x}_{[0,T]}$ based on cycle flows such that dynamical information from all timescales is contained in the new matrix. We can then use the MSM algorithm to obtain the desired fuzzy partition. This result will also lead to a generalization of the well known modularity function \cite{Newman04}, which will be sensitive to directional information. Finally, we will demonstrate the power of our method by analyzing an nonlinear time series of seismic data \cite{TimeSeriesIrreversibility2012,TimeSeriesIrreversibility2013}, offering a new
way to analyze irreversible real world processes.

\section{Method}
Given the time series $\mathbf{x}_{[1,T]}$ and the partition $\{S_1,\ldots,S_N\}$, define the series of symbols $\mathbf{s}_{[1,T]}$ by setting $s_i = x$ iff $x_i\in S_x$. Define the counts $N_T(x) = \sum_{i=1}^T\delta(s_i=x)$ and $N_T(x,y) = \sum_{i=1}^{T-1} \delta(s_i=x, s_{i+1} = y)$ and recall that the maximum likelihood estimator of (\ref{eq:Tmatrix}) is given by
\begin{equation}
P_{xy} = \frac{N_T(x,y)}{N_T(x)}.
\end{equation}
Let $G=(V,E)$ be the weighted and directed recurrence network representing $P$. We assume $P$ to be ergodic \footnote{One can always force $P$ to be ergodic by adding a small teleportation probability \cite{Lambiotte2012tel}. Here, ergodicity can be guaranteed by connecting the vertex last visited with the vertex first visited.} and thus $G$ to be strongly connected, such that the invariant distribution $\pi$ of $P$ exists and is unique.\\
Our aim is to construct an undirected graph that captures information flow in $G$, based on a cycle decomposition of the probability flow governed by $P$. More precisely, we will use the idea of counting cycles to count recurrences between nodes and capture the amount of communication between nodes given by the data. In terms of network modules, using cycle flows will account for considering multi-step, bidirectional connections between nodes which can reveal modular structure consisting of nodes communicating in both directions via short paths.\\
We now briefly introduce the theory of cycle decompositions for Markov chains as developed in \cite{Kalpazidou2006} and \cite{Qian2004}. An $n$-cycle on $G$ is an ordered sequence \footnote{More precisely, cycles are equivalence classes of ordered sequences up to cyclic permutations. In this note we do not distinguish between cycles and their representatives.} of $n$ connected nodes $\gamma = (x_1,x_2,\ldots, x_n)$, whose length we denote by $|\gamma| = n$. We consider the collection $\mathcal{C}$ of simple cycles on $G$, where no self-intersections are allowed. We proceed by describing an algorithm that generates counts $N_T^\gamma$ for every $\gamma\in\mathcal{C}$ based on counting recurrences along $\mathbf{s}_{[1,T]}$ \cite{Qian2004}. 
Let $t$ be the earliest time the recurrence $s_{t'} = s_t$ happens for some $t'<t$, i.e. the first time a node visited in the past has been revisited. The sequence $\mathbf{s}_{[t',t]}$ forms a simple cycle $\gamma$, so we increment $N_T^\gamma$ by one, exclude $\mathbf{s}_{[t',t]}$ from $\mathbf{s}_{[1,T]}$ and iterate the procedure, i.e. look for the next earliest recurrence along the remaining sequence, and so on. Then the limit
\begin{equation}
w(\gamma) := \lim_{T\rightarrow\infty} \frac{N_T^\gamma}{T}
\label{eq:WC}
\end{equation}
exists almost surely \cite{Qian2004} and gives us a uniquely defined probabilistic {\em cycle decomposition}, that is a collection $\Gamma = \{\gamma\in\mathcal{C}|w(\gamma)>0\}$ of cycles with positive weights $w(\gamma)$ such that for every edge $(xy)\in E$ the flow decomposition formula holds:
\begin{equation}
F_{xy} = \sum_{\gamma\supset(xy)} w(\gamma)
\label{eq:flowdecomp}
\end{equation}
where $F_{xy} = \pi_xP_{xy}$ is the probability flow through $(xy)$ and we write $\gamma\supset (xy)$ if the edge $(xy)$ is in $\gamma$. An explicit but computationally impractical formula to calculate the weights $w(\gamma)$ directly from $P$ was given in \cite{Qian2004}. 

\subsection{Example: The barbell graph} 
\begin{figure}[!ht]
    \centering
    \includegraphics[width=0.4\textwidth]{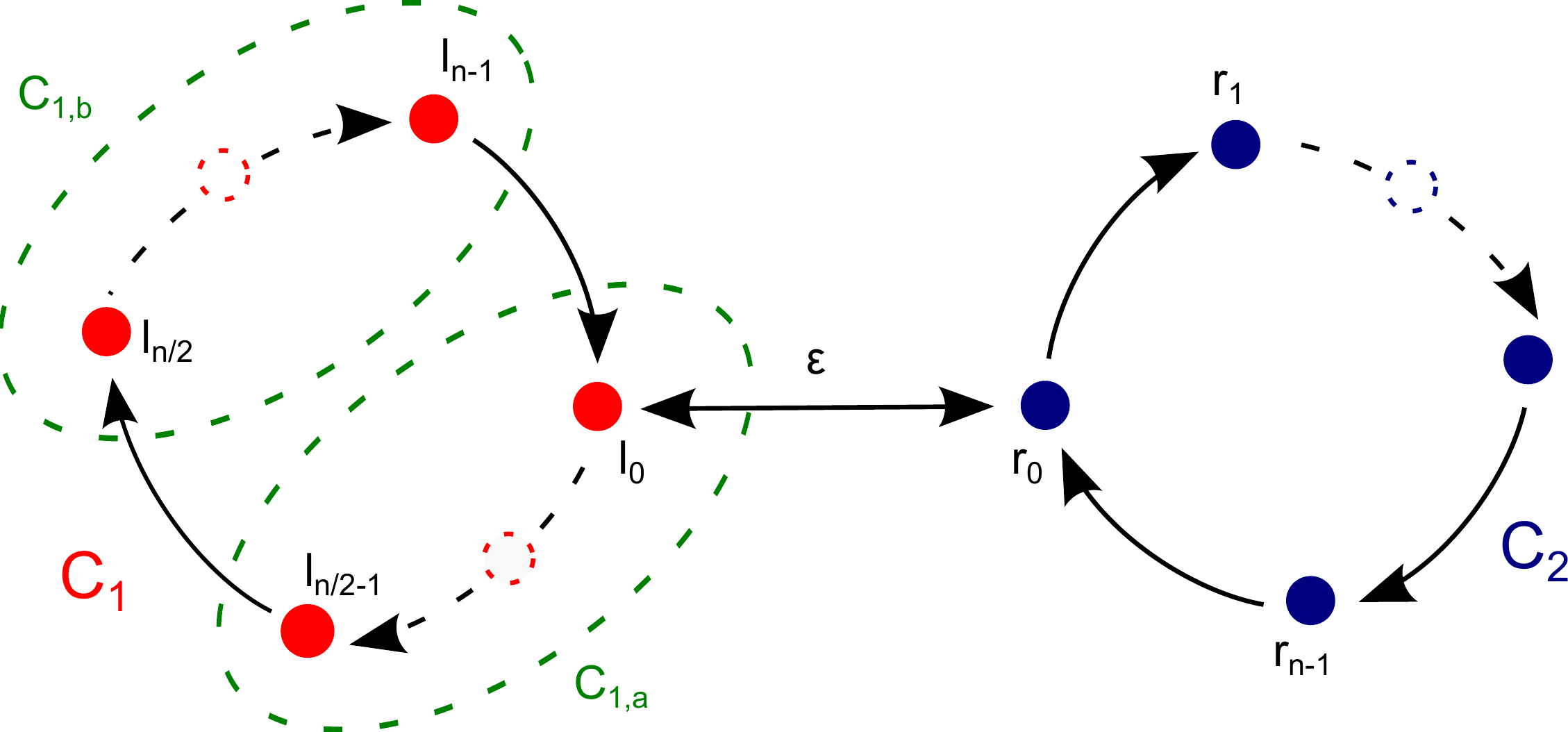}
    \caption{The barbell graph}
    \label{fig:barbell}
\end{figure}

As an example consider the unweighted barbell graph consisting of two cycles with $n$ nodes each, presented in Figure \ref{fig:barbell}. Since every edge belongs to exactly one of the three cycles $\alpha_c = (l_0, r_0)$, $\alpha_l = (l_0, l_1, \ldots, l_{n-1})$ and $\alpha_r = (r_0, r_1,\ldots, r_{n-1})$, the weights of these cycles can be inferred directly from (\ref{eq:flowdecomp}):
\begin{equation}
w(\alpha_l) = w(\alpha_r) = \frac{1}{2(n+1)}=: w, \quad w(\alpha_c) = w.
\end{equation}

We can use the idea of counting cycles via recurrences of $\mathbf{s}_{[1,T]}$ to count recurrences between any two nodes $x,y\in V$. We count every time $\mathbf{s}_{[0,T]}$ completes a cycle $\gamma$ as one recurrence for every $x,y\in \gamma$, and we normalize by $1/|\gamma|$ to account for the fact that longer cycles reflect less communication between $x$ and $y$. This leads to
\begin{equation}
 \tilde N_T(x, y) = \sum_{\gamma \ni x,y} \frac{1}{|\gamma|} N_T^\gamma
 \label{eq:counts}
\end{equation}
for the normalized number of recurrences between $x$ and $y$. Defined this way, $\tilde N_T(x,y)$ acts as a measure for the amount of communication between $x$ and $y$ seen by the data $\mathbf{s}_{[1,T]}$. The normalization is such that
\begin{equation}
\sum_y \tilde N_T(x,y) = \sum_{\gamma \ni x} N_T^\gamma = N_T(x).\label{eq:norm}
\end{equation}
The last equation is true because every return to $x$ corresponds to exactly one recurrence through a cycle $\gamma$ containing $x$.
Finally, we define the {\em communication intensity} $I_{xy}$ by passing to the limit of infinite observational time:
\begin{equation}\label{eq:I}
 I_{xy} := \lim_{T\rightarrow \infty} \frac{\tilde N_T(x,y)}{T} = \sum_{\gamma \ni x,y} \frac{w(\gamma)}{|\gamma|},
\end{equation}
using (\ref{eq:WC}) and (\ref{eq:counts}), and we arrive at the probabilistic cycle decomposition introduced earlier. Intuitively, $I_{xy}$ is large if there are many cycles connecting $x$ and $y$, and if they are important ($w(\gamma)$ large) and short ($|\gamma|$ small). The normalization (\ref{eq:norm}) directly translates into $\sum_y I_{xy} = \pi_x$. This allows us to introduce the main result of the paper, the {\em cycle transition matrix} $\mathcal{P}$ with components
\begin{equation}
 \mathcal{P}_{xy} = \lim_{T\rightarrow \infty} \frac{\tilde N_T(x,y)}{N_T(x)} = \frac{I_{xy}}{\pi_x} = \frac{1}{\pi_x}\sum_{\gamma\ni x,y} \frac{w(\gamma)}{|\gamma|}.
 \label{eq:CMSM}
\end{equation}
Note that $\mathcal{P}$ is reversible since $I_{xy} = I_{yx}$ and that it has the same stationary distribution as $P$, namely $\pi$. Counting cyclic recurrences has provided us with a way to symmetrize $P$, where the directional information is encoded in the sum over cycles of all lengths in (\ref{eq:CMSM}). Moreover, $I_{xy}$ gives us a transformation of $G$ into an undirected, weighted network $G_U$, where we connect two nodes $x$, $y$ by an edge with weight $I_{xy}$ if $I_{xy} > 0$. We will refer to $G_U$ as a \textbf{communication graph}. In \cite{Djurdjevac14} we discuss the properties of $\mathcal{P}$ and $I$ in more detail.

\section{Algorithm and computational complexity}

\subsection{Estimating $\mathcal{P}$}

Given the data $\mathbf{s}_{[0,T]}$, an estimator for the cycle transition matrix $\mathcal{P}$ is obtained by normalizing the counts $\tilde N_T(x,y)$ via (\ref{eq:CMSM}). The computation of these counts is done by the following algorithm, in view of (\ref{eq:counts}):
\begin{enumerate}
 \item [(i)] Initialization: set all $\tilde N_T(x,y) = 0$.
 \item [(ii)] Find the earliest $t$ such that $s_{t'} = s_{t}$ for some $t' < t$. Set $\gamma = \mathbf{s}_{[t',t]}$.
 \item [(iii)] Update $\tilde N_T(x,y) \rightarrow \tilde N_T(x,y) + \frac{1}{|\gamma|}$ for all $x,y \in \gamma$. Remove $\mathbf{s}_{[t',t]}$ from $\mathbf{s}_{[0,T]}$, and go back to (ii).
\end{enumerate}
The algorithm terminates when all recurrences in the data are removed. The remaining path can either be neglected or treated as another cycle, which is equivalent to connecting $s_T$ with $s_0$. The impact of this choice diminishes with large $T$. The algorithm is $\mathcal{O}(T)$ and not significantly more expensive then the computation of the counts $N_T(x,y)$ for the estimator of $P$, see (\ref{eq:Tmatrix}).

\subsection{Clustering}

With the estimator of $\mathcal{P}$ and the undirected network $G_U$ at our hands, we can, at least in principle, use any clustering method designed for undirected networks to partition $G_U$. Some methods might be more suitable then others, depending on additional properties of the data $\mathbf{s}_{[0,T]}$, and thus can be chosen on a case-by-case basis. In this paper we will use the MSM method \cite{Djurdjevac2011,SarichNet11} to cluster $G_U$ and refer to the whole algorithm as \emph{cycle MSM (CMSM)}. The main reason for this choice is that MSM is a dynamics-based method which can find multiscale fuzzy clusters.
More precisely, MSM clustering identifies modules $C_1,\ldots, C_m$ as the metastable sets of the random walk process on $G_U$ which has $\mathcal{P}$ as its transition matrix \cite{Djurdjevac14}. It also identifies a transition region $T=V\setminus(\bigcup_{i=1}^m C_i)$, which is not clustered and whose size can be tuned with a resolution parameter $\alpha$ \cite{SarichNet11}. Fuzzy affiliation functions are obtained as
\begin{equation}
q_i(x) = \pP(X_t\:\mbox{hits $C_i$ next}|X_0 = x), \quad \forall x\in V,
\end{equation}
by solving sparse, symmetric and positive definite linear systems \cite{MeSchEve,Djurdjevac2010}. It is easy see that $q_1,\ldots, q_m$ form a partition of unity $\sum_{i=1}^m q_i(x)=1\: \forall x\in V$, such that we can interpret $q_i(x)$ as the natural random walk based probability of affiliation of a node $x$ to a module $C_i$.\\
\textbf{Remark: }We can also use the CMSM algorithm in the case where only the network $G$ and no time series $\mathbf{s}_{[0,T]}$ is given. In such a case, we define a random walk process on the network. There are many ways of doing that, the simplest one is to obtain a transition matrix $\hat P$ by normalizing the edge weights. $\hat P$ is then used to generate a sample $\mathbf{\hat s}_{[0,T]}$ which serves as input for the CMSM algorithm. Unlike the time series case, the sampling time $T$ is not given a priori, it must be chosen by the user instead. How large $T$ has to be in order to obtain good convergence depends on the slowest relaxation timescale and hence on the second largest eigenvalue of $\hat P$ \cite{Sarich2010msm}. If $\hat P$ is very metastable, then sampling can become prohibitively expensive, and alternative ways to estimate $\mathcal{P}$ must be sought, which is beyond the scope of this article.

\section{Comparing CMSM with other methods}

Different clustering methods are based on different principles depending on the assumptions what the network actually represents. As discussed above, many existing approaches for clustering directed networks are based on using probability and information flow. Therefore, it is interesting to compare our CMSM method to these approaches, represented here by Infomap \cite{Delvenne2010}, Markov stability \cite{Lambiotte09} and modularity optimization \cite{Newman2008}.\\
Infomap \cite{Delvenne2010} is a popular method for detecting communities, relying on the idea that community structure can be used to describe the position of a random walker on the network compactly by reusing codewords in different communities. Infomap can deal with directed networks and performed very well in a recent benchmark \cite{Fortunato2009} with clique-like communities, but despite its information-theoretic origin it is inherently a one-step method and as such it can fail at detecting non-clique-like communities \cite{Schaub2012} by displaying an overpartitioning effect.\\
Markov stability (MS) is another state-of-the-art approach for community detection, which is based on revealing communities at different scales by looking at how the probability flow spreads out over time. At the heart of MS is the optimization of the stability function
\begin{equation}
\label{eq:stability}
r(t) = \mbox{trace}\: H^T[\Pi P(t) - \pi^T\pi]H
\end{equation}
where $\Pi = \mbox{diag}(\pi)$, $P(t) = P^t$ is the $t$-step transition matrix\footnote{In the case of continuous time, $P(t) = \exp(tL)$.} and $H$ is an indicator matrix encoding the community assignments of the nodes. For any fixed $t$, $r(t)$ encodes information about paths of length $t$, and the method uses $t$ as a resolution parameter: The optimization of $r(t)$ is carried out for all values of $t$ in the desired range, and one searches for communities which persist for a range of values of $t$. The well-known modularity function $Q$ \cite{Newman04,NewmanGirvan1} fits into the MS framework since $Q = r(1)$ \cite{Delvenne2010}. However unlike MS, modularity is a one-step method and has the same limitations as Infomap when faced with non-clique like communities which may appear in real networks \cite{Schaub2012}.\\
In contrast, at the heart of our approach is the matrix $\Pi\mathcal{P}$, which contains information about \textbf{all} cyclic paths. To illustrate this, we define a modified modularity function
\begin{equation}
 \bar Q = \mbox{trace}\: H^T[\Pi \mathcal{P} - \pi^T\pi]H.
 \label{eq:modularity_new}
\end{equation}
The difference between $r(t)$ and $\bar Q$ is the following: $r(t)$ finds communities with the property that returning to the community one started in after exactly $t$ steps is high. $\bar Q$ finds communities with the property that if one starts within one community, say at node $x$, selects a cycle $\gamma \ni x$ (at random according to the distribution $w(\gamma)$) and an exit node $y\in \gamma$ (at random with uniform probability), then the probability that $y$ is in the same community as the starting node $x$ is high.
Thus $\bar Q$ contains information about all cyclic paths at once, and cycles of length $t$ are discounted with a factor of $1/t$ in (\ref{eq:CMSM}).\\
There is no free parameter in the construction of $\mathcal{P}$, but during the MSM clustering stage $\alpha$ plays the same role as $t$ in MS \cite{SarichNet11}. Indeed the richness of MS and MSM clustering lies in the freedom of choosing a resolution parameter, but MS is limited to hard clustering, while MSM clustering is not. The key contribution of this paper is that it makes methods like MSM clustering available in situations where $P$ is irreversible.\\
We proceed by discussing the similarities and differences of the methods mentioned above and CMSM on two illustrative examples, where the time series for CMSM is sampled from a random walk process.

\subsection{The barbell graph (continued)}
\begin{figure}[!ht]
    \centering
    \includegraphics[width=0.4\textwidth]{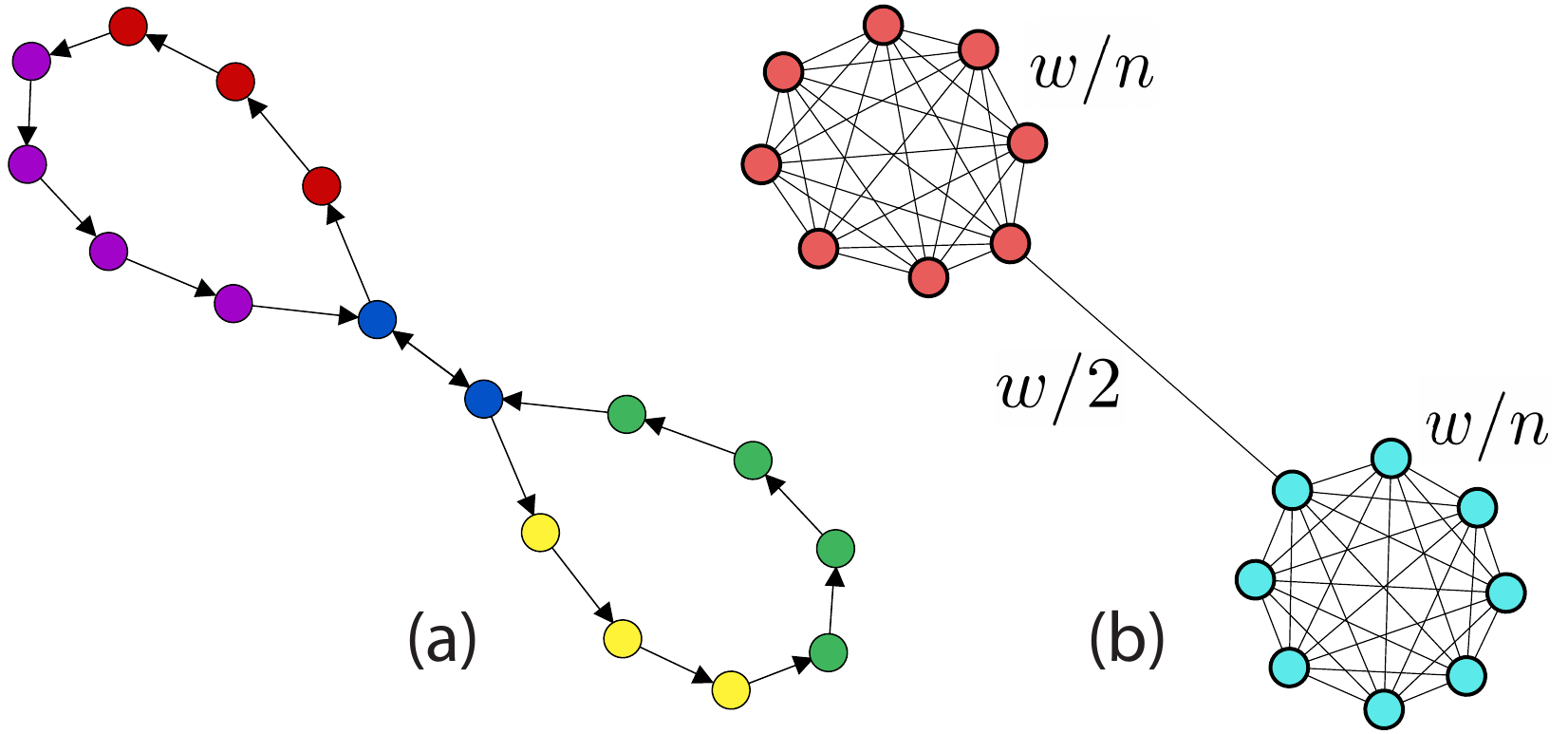}
    \caption{The barbell graph: (a) Clustering produced by Infomap. (b) Communication graph of the barbell graph and clustering produced by CMSM and MS algorithms.}
    \label{fig:barbellCompare}
\end{figure}

The barbell graph from Figure \ref{fig:barbell} is an example network with non clique like modules, which are often appearing in geographical, transport and distribution networks. After calculating the node communication intensity $I_{xy}$ explicitly, we obtain the undirected, weighted communication graph $G_U$ shown in Figure \ref{fig:barbellCompare} (b). By doing this, cycles $\alpha_l$ and $\alpha_r$ were mapped into two complete subgraphs with equally weighted edges, which resulted in the appearance of two modules $C_1 = \alpha_l$ and $C_2 = \alpha_r$. The same clustering is found by MS as the most stable partition. This is not surprising since MS was shown to be successful in recovering non-clique-like modules previously \cite{Schaub2012}. However, modularity optimization and Infomap face with the problem of overpartitioning of cycles. To demonstrate this problem, let us compare the modularity score $Q_1$ of a partition into $C_1, C_2$ with the modularity score $Q_2$ when $C_1$ is split into
two chains of equal size $C_{1,a}$ and $C_{1,b}$, see Figure \ref{fig:barbell}. The total change $\Delta Q$ in $Q$ under this split is $\Delta Q = \frac{1}{8} - \frac{1}{n+1}$. Thus, $\Delta Q >0$ for $n\geq 7$, such that as $n$ grows $Q$ favors a partition into more and more subchains with less then $7$ nodes over the partition $C_1,C_2$, even though increasing $n$ actually increases the metastability of the partition $C_1,C_2$. In contrast, the small chains favored by $Q$ are not metastable at all. Infomap faces similar problems and produces the partition shown in Figure \ref{fig:barbellCompare} (a), for a more detailed discussion see \cite{Fortunato2009,Schaub2012}.

\subsection{A network with directed (non-)modular structure}

\begin{figure}[!ht]
    \centering
    \includegraphics[width=0.48\textwidth]{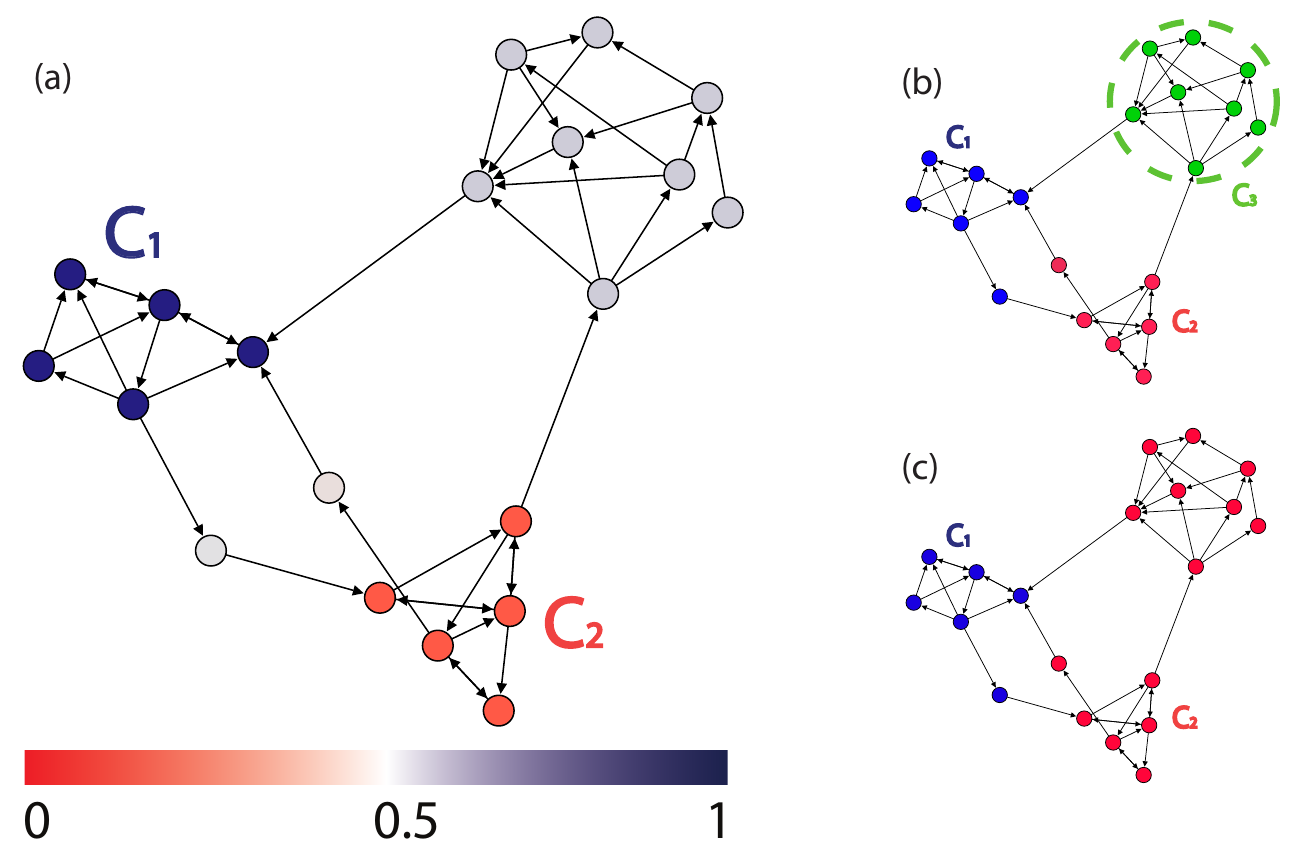}
    \caption{Example network with directed (non-)modular structure.(a) Fuzzy clustering produced by CMSM. Clustering produced by (b) modularity optimization and Infomap, (c) MS method.}
    \label{fig:marcoex}
\end{figure}
The next example is a network with $21$ nodes, for which CMSM clustering finds two metastable modules: $C_1$ colored in blue and $C_2$ colored in red, see Figure \ref{fig:marcoex}(a). The rest of the network forms a large transition region consisting of nodes with affiliation probability less than $0.8$. If we cluster this network using the modularity optimization (or Infomap) algorithm, a third module $C_3$ appears (green in \ref{fig:marcoex}(b)). However, $C_3$ is not a metastable module because none of its nodes are connected via short paths in both directions. For example, $A, B\in C_3$ are connected by a directed edge $(AB)$, but in the direction from $B$ to $A$ they are connected only by long paths that pass through the whole network. Consequently, $I_{AB}$ is small and therefore CMSM-clustering overcomes this problem, improving upon existing one-directional density based methods. The only stable partition found by MS is shown in Figure \ref{fig:marcoex} (c). Due to the benefit of
looking at walks of different length, MS recognizes the full structure of the network and obtains two modules,
but because it can produce only hard partitions nodes from $C_3$ get assigned to $C_2$.

\section{A time series of earthquakes}
Recurrence networks are frequently used to analyse seismic data. See \cite{Abe2006, Abe2004, Davidsen2008,TimeSeriesIrreversibility2013} for a discussion of several approaches, including the one based on partitioning $\Omega$ used in this article. Our final example is thus a time series $\{x_1,\ldots,x_T\}$ of seismic events in California from 1952 to 2012, obtained from the SCEC\footnote{Southern California Earthquake Center, \url{www.scec.org}}. Only events with magnitude larger than $m_c = 2.5$ are considered (these are $48669$ events). The observational space $\Omega$ is the rectangle from $32^\circ$ to $37^\circ$ in latitude and $-122^\circ$ to $-114^\circ$ in longitude, and we partition $\Omega$ into $4000$ quadratic boxes $S_i$ of length $\Delta l = 0.1^\circ$. Finally, the boxes which don't see any events are discarded. The transition matrix (\ref{eq:Tmatrix}) thus
constructed corresponds to a network with $2175$ nodes and $28839$ edges.\\
\begin{figure*}[ht]
    \centering
      \includegraphics[width=0.82\textwidth]{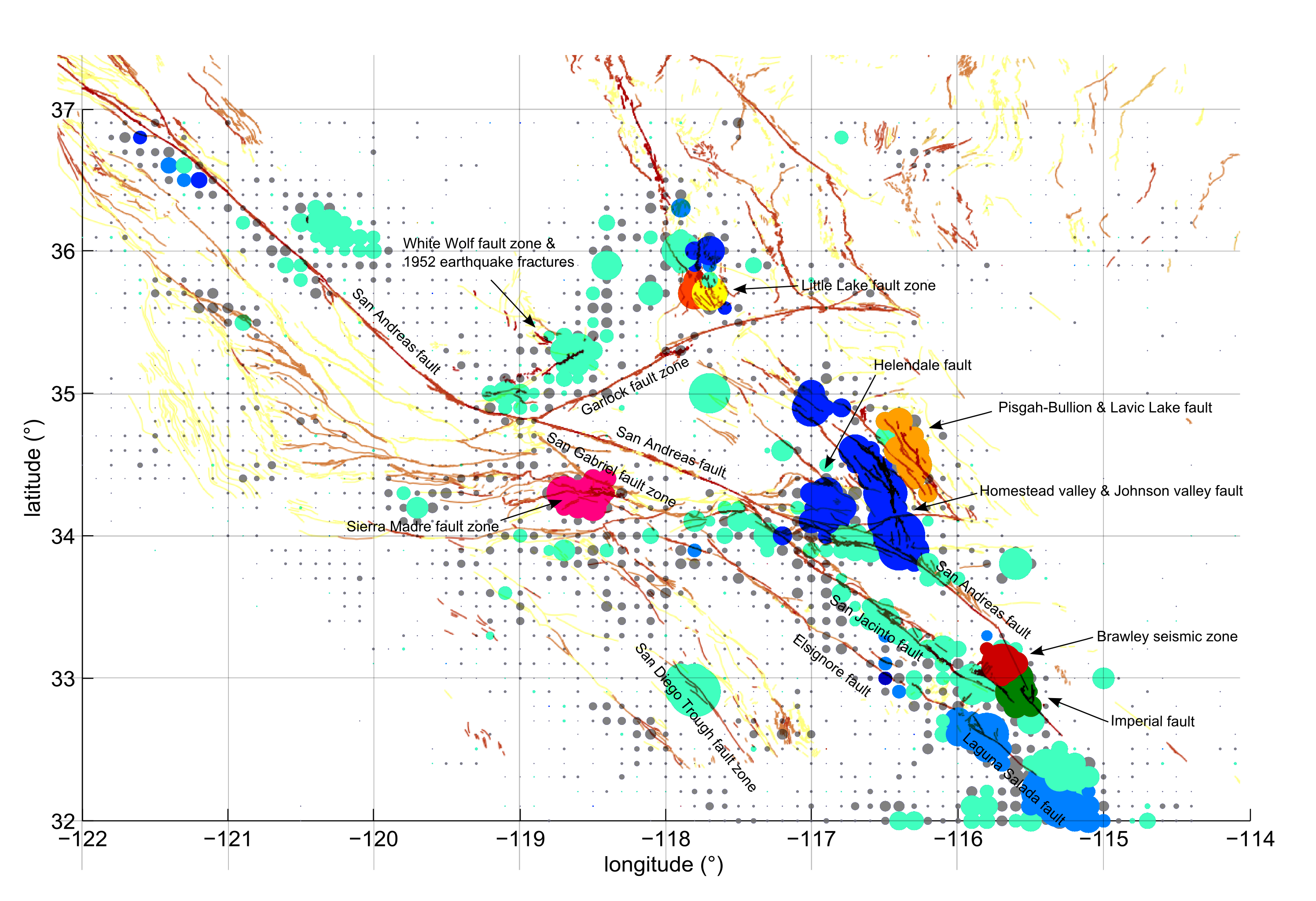}
   \caption{Quaternary faults \cite{USGS} in Southern California and the clustering of the SCEC time series found by CMSM. Node size is proportional to the number of events, color indicates the modules found.}
    \label{fig:EarthquakeNetRes}
\end{figure*}
The CMSM algorithm implemented in \texttt{matlab} constructs the estimator of $\mathcal{P}$ and $G_U$ in $2.05$ seconds on a laptop,  clusters $G_U$ in $7.6$ seconds and reports $7739$ cycles. This clearly shows that performance is not an issue when the CMSM algorithm is used on time series data. The fuzzy clustering obtained by CMSM is shown in Figure \ref{fig:EarthquakeNetRes}, where a node $x$ receives the color of module $C_i$ if $q_i(x) \geq 0.8$, and is colored grey if $q_i(x) < 0.8$ for all modules $C_i$. In fact the latter is the case for $80\%$ of the nodes, but these correspond to only $25\%$ of all events. This illustrates that our fuzzy clustering correctly reflects the uncertainty coming from limited data. A full clustering obtained by e.g. MS or Infomap would have to cluster the grey nodes as well, even though not enough data is available to do so. CMSM-clustering finds 9 modules, all of which correspond to important faults or
groups of faults, the largest one containing the San Andreas fault. This demonstrates that our method can successfully uncover structure in the dataset $\mathbf{x}_{[0,T]}$ - in this case, the presence of geological faults that influence the earthquake pattern.

\section{Conclusion}
In this paper we addressed the problem of module detection in weighted directed networks coming from time series data. The new method we propose is based on constructing a reversible transition matrix $\mathcal{P}$ which is based on using multi-step, bidirectional transitions encoded by a cycle decomposition of the probability flow, and we provide a simple and fast algorithm for estimating $\mathcal{P}$ directly from the timeseries data. Since $\mathcal{P}$ is reversible, it allows us to apply clustering methods designed for undirected graphs. We applied the method to several examples and showed how it overcomes essential limitations of common methods. Finally we demonstrated our novel approach on a real-world directed network coming from time-series data, offering a new way to analyze irreversible processes.

\acknowledgments
The authors thank Stefan R\"udrich for valuable insights on earthquake data analysis and useful feedback on the manuscript; and Christof Sch\"utte, Marco Sarich and Michael Schaub for helpful discussions. The authors further thank the two anonymous referees for comments that improved the paper.

\bibliographystyle{plain}
\bibliography{eplpaper}
\end{document}